\documentstyle[12pt]{article}
\begin{document}
\newcommand {\ber} {\begin{eqnarray*}}
\newcommand {\eer} {\end{eqnarray*}}
\newcommand {\bea} {\begin{eqnarray}}
\newcommand {\eea} {\end{eqnarray}}
\newcommand {\beq} {\begin{equation}}
\newcommand {\eeq} {\end{equation}}
\newcommand {\state} [1] {\mid \! \! {#1} \rangleg}
\newcommand {\sym} {$SY\! M_2\ $}
\newcommand {\eqref} [1] {(\ref {#1})}
\newcommand{\preprint}[1]{\begin{table}[t] 
           \begin{flushright}               
           \begin{large}{#1}\end{large}     
           \end{flushright}                 
           \end{table}}                     
\def\Acknowledgements{\bigskip  \bigskip {\begin{center} \begin{large}
             \bf ACKNOWLEDGMENTS \end{large}\end{center}}}

\newcommand{\half} {{1\over {\sqrt2}}}
\newcommand{\dx} {\partial _1}

\def\Dslash{\not{\hbox{\kern-4pt $D$}}}
\def\cmp#1{{\it Comm. Math. Phys.} {\bf #1}}
\def\cqg#1{{\it Class. Quantum Grav.} {\bf #1}}
\def\pl#1{{\it Phys. Lett.} {\bf #1B}}
\def\prl#1{{\it Phys. Rev. Lett.} {\bf #1}}
\def\prd#1{{\it Phys. Rev.} {\bf D#1}}
\def\prr#1{{\it Phys. Rev.} {\bf #1}}
\def\pr#1{{\it Phys. Rept.} {\bf #1}}
\def\np#1{{\it Nucl. Phys.} {\bf B#1}}
\def\ncim#1{{\it Nuovo Cimento} {\bf #1}}
\def\lnc#1{{\it Lett. Nuovo Cim.} {\bf #1}}
\def\jmath#1{{\it J. Math. Phys.} {\bf #1}}
\def\mpl#1{{\it Mod. Phys. Lett.}{\bf A#1}}
\def\jmp#1{{\it J. Mod. Phys.}{\bf A#1}}
\def\aop#1{{\it Ann. Phys.} {\bf #1}}
\def\hpa#1{{\it Helv.Phys.Acta} {\bf#1}}
\def\mycomm#1{\hfill\break{\tt #1}\hfill\break}
\def\dslash{\not{\hbox{\kern-4pt $\partial$}}}
\begin{titlepage}
\rightline{TAUP-2509-98}
\rightline{WIS-98/19/July-DPP}
\rightline{\today}
\vskip 1cm
\centerline{{\Large \bf  Screening in Supersymmetric Gauge Theories}}
\centerline{{\Large \bf in Two Dimensions}}
\vskip 1cm
\centerline{A. Armoni$^a$
\footnote{e-mail: armoni@post.tau.ac.il}
, Y. Frishman$^b$
\footnote{e-mail: fnfrishm@wicc.weizmann.ac.il}
 and J. Sonnenschein$^a$
\footnote{e-mail: cobi@post.tau.ac.il}
\footnote{work supported in part by the Israel Science Foundation, the US-Israel Binational
Science Foundation and the Einstein Center for Theoretical Physics at
the Weizmann Institute. }}
\vskip 1cm
\begin{center}
\em $^a$School of Physics and Astronomy
\\Beverly and Raymond Sackler Faculty of Exact Sciences
\\Tel Aviv University, Ramat Aviv, 69978, Israel
\\and
\\ $^b$Department of Particle Physics
\\Weizmann Institute of Science
\\76100 Rehovot, Israel
\end{center}
\vskip 1cm

\begin{abstract}
 We show that the string tension in ${\cal N}=1$ two-dimensional super
 Yang-Mills theory vanishes
independently of the representation of the quark anti-quark external
 source. We argue that this result persists in $SQCD _2$ and in
two-dimensional gauge theories with extended supersymmetry or in
chiral invariant models with at least one massless dynamical fermion.
 
We also compute the string tension for the massive Schwinger model, as a
demonstration of the method of the calculation.
\end{abstract}
\end{titlepage}

\section{Introduction} 
 In a paper by Gross et al.\cite{gross} it was conjectured that
 two dimensional  ${\cal N}=1$ Super Yang-Mills exhibits a screening
 nature for any representation of the external sources. In particular
 the adjoint vector multiplet can screen completely an external
 source which transforms in the fundamental representation.

\sym is an interesting theory, since it is probably the simplest
non-trivial supersymmetric model in $1+1$ dimensions. It's dynamics is
similar to the adjoint fermions model, but SUSY simplifies its
behavior. The model was analyzed lately in\cite{MSS,ALP}.

In this letter we present a short proof that the theory is
indeed screening. We show that that the string tension vanishes in this
 model (though we don't exclude less than linear confining potential). An evidence to this phenomenon was given
in\cite{AS}. The proof follows an idea presented in\cite{CJS} which
was generalized to the non-Abelian case in\cite{AFS}. We use a local chiral
rotation to eliminate the external source from the action. The chiral
rotation affects terms which are not chiral invariant. In the
case of \sym it is the interaction term of the gluino and the
pseudo-scalar. The string tension is then computed using the change in
the Hamiltonian density.

As a demonstration of the technique, we compute the string tension in
the massive Schwinger model. A different
derivation in the fermionic basis was given in\cite{Adam}. 

The expression for the string tension in the massive Schwinger model 
was calculated by using bosonization\cite{CJS}
\beq
\sigma _{QED} = m\mu \left (1-\cos (2\pi {q_{ext} \over q_{dyn}})
\right ) +O(m^2) \label{abelian},
\eeq
where $m$ is the electron mass, $\mu = g{\exp (\gamma) \over 2\pi^{3/2}}$
($g$ is the gauge coupling and $\gamma$ is the Euler number)
 and $q_{ext}$,$q_{dyn}$ are the external and dynamical charges
respectively. 

The expression in massive $QCD_2$ is\cite{AFS}
\beq
\sigma _{QCD} =  m \mu _R \
\sum _i \left ( 1-\cos (4\pi \lambda _i  {k_{ext}  \over k_{dyn}}
  )\right ) +O(m^2) \label{nonabelian}
\eeq
where $\mu _R \sim g$, $\lambda _i$ are the isospin eigenvalues of the
dynamical representation, $k_{ext}$ and $k_{dyn}$ are the affine
current algebra levels of the external and dynamical representations, respectively. This
expression is valid only for the fundamental and the adjoint
representations. Other representations were also discussed
in\cite{AFS}.

Note that when $m=0$ the string tensions \eqref{abelian} and
\eqref{nonabelian} vanish. In this case the theories are screening due
to chiral invariance of the actions.
 
The $O(m^2)$ term in \eqref{abelian} and  \eqref{nonabelian} indicates
that these expressions are only the leading terms in mass perturbation
theory and are valid when $m\ll g$. The next to leading order
correction, in the Abelian case, was derived in\cite{Adam}. 

The derivation of the string tension in the massive Schwinger model is
as follows. Consider the partition function of two dimensional massive QED
\bea 
\lefteqn{Z=} \label{partition} \\ &&
\int DA_\mu D\bar \Psi D \Psi \exp \left ( i\int d^2x\ (-{1\over 4g^2} F_{\mu\nu}^2
+\bar \Psi i\dslash \Psi - m\bar \Psi \Psi - q_{dyn}A_\mu \bar \Psi \gamma ^\mu
\Psi )\right ), \nonumber
\eea
where $q_{dyn}$ is the charge of the dynamical fermions.
Let us add an external electron-positron source with charge $q_{ext}$
at $\pm L$, namely $j_0^{ext} = q_{ext}(\delta (x+L) - \delta
(x-L))$, so that the change of ${\cal L}$ is $-j_\mu ^{ext} A^{\mu}
(x)$. Note that by choosing $j_\mu ^{ext}$ which is conserved,
$\partial ^\mu j^{ext}_\mu =0$, the action including the coupling to the
external current is gauge invariant.

Now, to eliminate this charge we perform a local, space-dependent
 left-handed rotation
\bea
 \Psi \rightarrow e^{i\alpha(x) {1\over 2}(1-\gamma_5) } \Psi \\
  \bar \Psi \rightarrow \bar \Psi e^{-i\alpha(x) {1\over 2}(1+\gamma_5) } , 
\eea
where $\gamma ^ 5 = \gamma ^0 \gamma ^1$.
The rotation introduce a change in the action, due to the chiral anomaly
\bea
\delta S = \int d^2 x {\alpha(x) q_{dyn} \over 4\pi} F,
\eea
where $F$ is the dual of the electric field $F={1\over 2} \epsilon
^{\mu \nu} F_{\mu \nu}$.

The new action of the original fields is
\bea \label{rotated}
\lefteqn{S=
\int d^2x\ (-{1\over 4g^2} F_{\mu\nu}^2
+\bar \Psi i\dslash \Psi -\bar \Psi \partial_\mu \alpha(x) \gamma ^\mu
{1\over 2}(1-\gamma _5) \Psi } \\ &&
 - m\bar \Psi e^{-i\alpha(x) \gamma_5 }\Psi
 - q_{dyn}A_\mu \bar \Psi \gamma ^\mu\Psi - q_{ext}(\delta (x+L) -
 \delta (x-L))A_0 +  {\alpha(x) q_{dyn} \over 4\pi} F) \nonumber
\eea

The external source and the anomaly term are similar, both being
linear in the gauge potential. This is the
reason that the $\theta$-vacuum and electron-positron pair at the boundaries are the same in
two-dimensions\cite{CJS}. In the following we assume $\theta =0$, as
otherwise we absorb it in $\alpha $. Choosing the $A_1=0$ gauge and integrating by
parts the anomaly term looks like an external source
\beq
{q_{dyn}\over 2\pi} A_0 \partial_1 \alpha(x)
\eeq 
This term can cancel the external source by the choice
\beq
  \alpha(x) = 2\pi {q_{ext} \over q_{dyn}} (\theta (x+L) -\theta
   (x-L)).
\eeq 
Let us take the limit $L\rightarrow \infty$.
The form of the action, in the region $B$ of $-L<x<L$ is
\beq
S_B =\int _B d^2x\ (-{1\over 4g^2} F_{\mu\nu}^2
+\bar \Psi i\dslash \Psi 
 - m\bar \Psi e^{-i 2\pi {q_{ext} \over q_{dyn}} \gamma_5 }\Psi
 - q_{dyn}A_\mu \bar \Psi \gamma ^\mu\Psi ) \label{rotated2}
\eeq
Thus the total impact of the external electron-positron pair is a chiral
rotation of the mass term. This term can be written as
\beq
\bar \Psi e^{-i 2\pi {q_{ext} \over q_{dyn}} \gamma_5 }\Psi=
\cos  (2\pi {q_{ext} \over q_{dyn}}) \bar \Psi \Psi -  i\sin (2\pi {q_{ext} \over q_{dyn}})
\bar \Psi \gamma_5 \Psi
\eeq
The string tension is the vacuum expectation value (v.e.v.) of the
Hamiltonian density in the presence of the external source relative to the
v.e.v. of the Hamiltonian density without the external source, in the
$L\rightarrow \infty$ limit.  
\beq
\sigma = <{\cal H}>-<{\cal H}_0>
\eeq
The change in the vacuum energy is due to the mass term. The change in
the kinetic term which appears in \eqref{rotated} does not contribute to the
vacuum energy\cite{AFS}.
Thus
\beq
\sigma = 
m\cos  (2\pi {q_{ext} \over q_{dyn}}) <\bar \Psi \Psi> -m \sin (2\pi {q_{ext} \over q_{dyn}})
<\bar \Psi i\gamma_5 \Psi> -m<\bar \Psi \Psi>
\eeq
Thus, the values of the condensates $<\bar \Psi \Psi>$  and $<\bar
\Psi \gamma_5 \Psi>$ are needed. The easiest way to compute these
condensates is Bosonization\cite{CJS}, but it can also be computed directly in
the fermionic language\cite{JSW}
\bea
 && <\bar \Psi \Psi>=  -g{\exp (\gamma) \over 2\pi^{3/2}}
 \label{condensate} \\
&& <\bar \Psi \gamma_5 \Psi>=0 ,\label{trivial}
\eea

The condensates \eqref{condensate} and \eqref{trivial} were computed in the massless
Schwinger model. However, the corrections to these expressions will
affect the string tension only by terms higher in ${m\over g}$.  
Eq.\eqref{trivial} is due to parity
invariance (with our choice $\theta =0$). The resulting string tension
is eq.\eqref{abelian}.

Though eq.\eqref{abelian} gives only the leading term in a $m/g$
expansion and might be corrected\cite{Adam}, when $q_{ext}$ is an integer multiple of
$q_{dyn}$ the string tension is {\em exactly} zero, since in this case
the rotated action\eqref{rotated2} is the same as\eqref{partition}.

\section{Super Yang-Mills}

The same technique can be used to prove screening in \sym.
In this case  the action is \cite{ferrara}  
\beq
 \label{sym} S = \int d^2 x \ tr \left ( -{1\over 4g^2} F^2_{\mu\nu} +  i\bar
 \lambda\Dslash\lambda
 +{1\over 2} (D_\mu \phi )^2 - 2ig \phi \bar \lambda \gamma _5 \lambda
\right ) ,
\eeq
where  $A_\mu $ is the
 gluon field, $\lambda $ the gluino (a Majorana fermion) and $\phi $
 a pseudo-scalar, are the components of the vector supermultiplet and
transform as the adjoint representation of $SU(N_c)$. Also $D_\mu =
\partial _\mu - i[A_\mu,.]$.

The action \eqref{sym} is invariant under SUSY
\ber
&&  \delta A_\mu = -i g\bar \epsilon \gamma _5 \gamma _\mu  \sqrt{2} \lambda \\
&&  \delta \phi = - \bar \epsilon \sqrt{2} \lambda \\
&&  \delta \lambda \ = {1\over {2 \sqrt{2}}g} \epsilon \epsilon ^{\mu \nu}
  F_{\mu \nu} +{i \over {\sqrt{2}}} \gamma ^{\mu} \epsilon D_{\mu} \phi
\eer

We now introduce an external current. The external
source breaks explicitly supersymmetry. However, this breaking does
not affect our derivation.
 We assume a semi-classical quark anti-quark
pair which points in some direction in the algebra. Without loss of
generality this direction can be chosen as the '3' direction
('isospin'). The additional part in the Lagrangian is $-tr\ j_\mu
^{ext} A^\mu$ where $j_0 ^{a\ ext} = [C(R_{ext})] \delta ^{a3}(\delta (x+L) - \delta (x-L))$
and $[C(R_{ext})]$ is  a c-number which depends on the representation
of the external source (see ref.\cite{AFS}). The interaction term can be eliminated by a left-handed
rotation in the '3' direction, of the gluino field (we are using a
spherical basis, and so we can perform appropriate complex
transformation also for real fermions)

\bea
 \lambda  \rightarrow \tilde \lambda = e^{i\alpha(x) {1\over 2}(1-\gamma_5 ) T^3} \lambda \\
  {\bar \lambda}  \rightarrow {\tilde {\bar \lambda }}  =
 \bar \lambda  e^{-i\alpha(x) {1\over 2}(1+\gamma_5 ) T^3} 
\eea
 $T^3$ is in the 3 direction of the adjoint representation 
\ber
 &&
T^3 = diag (\mu _1,\mu _2, ...,\mu _{N_c^2 -1}) \\
 &&       = diag (1,0,-1,
\underbrace{{1\over 2},-{1\over 2},{1\over 2},-{1\over 2},...,{1\over
    2},-{1\over 2}}_{2(N_c-2)\,\,\, doublets},\underbrace{0,0,...,0}_{(N_c-2)^2})
\eer

 The chiral
rotation introduces an anomaly term $tr\ {\alpha(x) T^3 \over 4\pi} F$,
which is used to cancel the external charges. Note that the
  chiral rotation introduces additional terms. However, these terms
 involve more derivatives and therefore do not
  affect the string tension. This situation is very similar to the
  Abelian case and $QCD_2$\cite{AFS}.  

The choice $\alpha (x) = 2\pi {C(R_{ext}) \over N_c} (\theta (x+L) -\theta
   (x-L))$ leads to an action which is similar to the
   original \eqref{sym}, but has a chiral rotated term. The
   information of the external source is now transformed into a rotation angle.

The terms which are relevant to the computation of the string tension
are those which appear in the interaction Lagrangian. In this case, it
is the gluino pseudo-scalar term
\beq
 tr \ 2i \phi \bar \lambda \gamma _5 \lambda \rightarrow 
 tr \ 2i \phi \tilde {\bar \lambda}  \gamma _5 {\tilde \lambda}
\eeq

Let us see how this change influences the Hamiltonian vacuum energy. In the
original theory, without the external source, $<H_0>=0$ since the theory is supersymmetric and
$H_0 \sim Q^2$ (where $Q$ is the supercharge). In particular it means that there is no
$<tr\ \phi {\bar \lambda}  \gamma _5 { \lambda}>$ condensate. Here we
assume that SUSY is not broken dynamically. The numerical analysis
of\cite{ALP} indicates that this is indeed the case.

Let us compute the Hamiltonian density of the rotated theory. In the
regime $-L < x < L$
\beq
<{\cal H}> = 2ig <tr\ \phi \tilde {\bar \lambda}  \gamma _5 {\tilde
  \lambda}>
\eeq
By using the fact that $T^3$ is diagonal, and the vacuum state is
color symmetric, we get
\bea
\lefteqn{ <tr\ \phi \tilde {\bar \lambda}  \gamma _5 {\tilde
    \lambda}>=} \\
&&
 {1\over {N_c^2-1}} \sum _a \cos (\alpha \mu _a) <tr\ \phi {\bar
   \lambda}  \gamma _5 { \lambda}>
-i {1\over {N_c^2-1}} \sum _a \sin (\alpha \mu _a) <tr\ \phi {\bar
   \lambda}  { \lambda}> \nonumber ,
\eea
where $\alpha = \lim _ {L \rightarrow \infty} \alpha (x)$.
The first term on the right hand side vanishes since as argued before $<tr\ \phi {\bar
  \lambda}  \gamma _5 { \lambda}>=0$, and the second term vanishes
since the isospin eigenvalues, $\mu _a$, come in pairs of opposite
signs.

Thus $<{\cal H}>=0$ and the string tension is zero. 

Note that though we used the classical expression for the external current
and the effective Hamiltonian may include other terms, these terms
cannot change the value of the string tension. It is so because this
theory contains only one dimension-full parameter, the gauge coupling
$g$, and therefore the string tension is some number times $g^2$. We
showed that this number is zero and higher terms in $g$
which may appear in the effective Hamiltonian cannot affect the string tension.

The meaning of the last result is that a quark anti-quark pair
 located at $x=\pm \infty$ does not generate a linear potential.
 In the non supersymmetric case, it is a consequence
of infinitely many adjoint fermions which are produced
from the vacuum, as there is no mass gap,  that are attracted to
the external source, form a soliton in the fundamental representation
 and result in screening it. We do not have a construction of the
 soliton, but such must occur as a result of the situation implied by
 the equivalence in\cite{KS}. We believe that similar mechanism occurs
 in the supersymmetric case too. A complementary argument\cite{FS,AS} is that due to
loop effects, the intermediate gauge boson acquires a mass $M^2 \sim g^2 N_c$, which
leads to a Yukawa potential between the external quark anti-quark pair. 

The above result can be generalized to theories with extended
supersymmetry and additional massive or massless matter content.

We argue that any supersymmetric gauge theory in two dimensions is
screening.
Technically, the reason is that the gluino is coupled to other fields
in such a way that $<{\cal H}>=0$ (guaranteed if SUSY is not
broken dynamically) and therefore there are no non-trivial chiral condensates. 
However, since the string tension is proportional to chiral
condensates, SUSY leads to zero string tension. 
Physically, it follows from the fact that the gluino is an adjoint
{\em massless} fermion. Since it does not acquire mass, external
sources are screened, as in the non-supersymmetric massless model.

In fact, the essential
requirement for a screening nature of the type argued above, is to
have among the charged particles at least one massless particle whose
masslessness is protected by an unbroken symmetry. The symmetry can be
gauge symmetry combined with supersymmetry or chiral symmetry.


\newpage
\begin{thebibliography}{99}

\bibitem{gross} D.J. Gross, I.R. Klebanov,A.V. Matytsin and
  A.V. Smilga,{\em ``Screening vs. Confinement in 1+1
    Dimensions''}, \np{461} (1996) 109.

\bibitem{MSS} Y. Matsumura, N. Sakai and T. Sakai,
{\em ``Mass Spectra of Supersymmetric Yang-Mills Theories in (1+1)
  Dimensions''}, \prd{52} (1995) 2446.

\bibitem{ALP} F. Antonuccio, O. Lunin and S. Pinsky,
{\em ``Nonperturbative Spectrum of Two-Dimensional (1,1) Super
  Yang-Mills at Finite and Large N''}, OHSTPY-HEP-TH-98-005,
hepth/9803170.

\bibitem{AS} A. Armoni and J. Sonnenschein,
{\em ``Screening and Confinement in large $N_f$ $QCD_2$ and in $N=1$
  \sym''}, \np{502} (1997) 516.

\bibitem{CJS} S. Coleman, R. Jackiw and L. Susskind,
{\em ``Charge Shielding and Quark Confinement in the Massive Schwinger
  Model''}, \aop{93} (1975) 267.

\bibitem{AFS} A. Armoni, Y. Frishman and J. Sonnenschein,
{\em ``The String Tension in Massive $QCD_2$''}, \prl{80} (1998) 430.


\bibitem{Adam} C. Adam,
{\em ``Charge screening and confinement in the massive Schwinger
  model''}, \pl{394} (1997) 161.

\bibitem{JSW} C. Jayewardena,
{\em ``Schwinger model on S(2)''},  \hpa{61} (1988) 633;
 I. Sachs and A.Wipf,
{\em ``Finite temperature Schwinger model''}, \hpa{65}
(1992) 652.

\bibitem{ferrara} S. Ferrara,{\em ``Supersymmetric Gauge Theories in
    two Dimensions''}, \lnc{13} (1975) 629.

\bibitem{KS} D. Kutasov and A. Schwimmer,
{\em "Universality in Two Dimensional Gauge Theory"}, \np{442} (1995) 447.

\bibitem{FS} Y. Frishman and J. Sonnenschein,
{\em ``QCD in two-dimensions, Screening, Confinement and Novel
  Non-Abelian solutions''}, \np{496} (1997) 285.



\end{thebibliography}
\end{document}